\documentclass[
aps,%
12pt,%
final,%
notitlepage,%
oneside,%
onecolumn,%
nobibnotes,%
nofootinbib,% In the current version of REVTeX, when this option is on,
superscriptaddress,%
noshowpacs,%
centertags]%
{revtex4}

\usepackage{rotating}

\usepackage{epsfig}
\usepackage{bm}% bold math

\def \ms {{\overline{\mbox{MS}}}}

\newcommand{\be}{\begin{equation}}
\newcommand{\ee}{\end{equation}}

\newcommand{\bea}{\begin{eqnarray}}
\newcommand{\eea}{\end{eqnarray}}

\begin{document}
%\selectlanguage{russian} % For a paper in Russian
%\selectlanguage{english} % For a paper in English
%
\title{$Q^2$-evolution of parton densities at small $x$
%low Q2
values.\\
Effective scale for combined H1$\&$ZEUS F2 data.}
\author{ A.V. Kotikov and B.G. Shaikhatdenov}
%\thanks{}
\affiliation{Joint Institute for Nuclear Research, Russia}
%%
%\author{\firstmame{...} \surname{...}}
%%\email[]{Your e-mail address}
%%\homepage[]{Your web page}
%%\thanks{}
%%\altaffiliation{}
\noaffiliation
\begin{abstract}
We use
%the Grunberg-like effective approach for the singular part of
%anomalous dimesnions to fix the scale of the coupling constant
%at the next-to-leading order of approximation
the Bessel-inspired behavior of the structure function $F_2$ at small $x$,
obtained for a flat initial condition in the DGLAP evolution equations.
We fix  the scale of the coupling constant, which eliminates
%is responsible for the vanishing o
%with the Grunberg-like effective approach for
the singular part of
anomalous dimesnions
%to fix the scale of the coupling constant
at the next-to-leading order of approximation.
The approach together with the ``frozen'' and analytic modifications of the
strong coupling constant is used to study the precise
combined H1$\&$ZEUS data for the structure function $F_2$
published recently.
\end{abstract}
\maketitle
%

%\section{Introduction}

%{\bf I}.
A standard way to study the $x$ behavior of
quarks and gluons is to compare the data
with the numerical solution to the
%DGLAP
Dokshitzer-Gribov-Lipatov-Altarelli-Parisi (DGLAP)
equations
\cite{DGLAP}
by fitting the parameters of
$x$-profile of partons at some initial $Q_0^2$ and
the
%QCD
energy scale $\Lambda$ of
Quantum Chromodynamics (QCD) \cite{fits,Ourfits}.
However, for the purpose of analyzing exclusively the
small-$x$ region, there is an alternative to carry out
%of doing
a simpler analysis
by using some of the existing analytical solutions to DGLAP equations
in the small-$x$ limit \cite{BF1,Q2evo}.

A reasonable agreement between H1 and ZEUS data  \cite{H1ZEUS}
%\cite{H1ZEUS,H1slo,DIS02}
and the next-to-leading-order (NLO) approximation of
perturbative QCD
%Quantum Chromodynamics (QCD)
has been observed for $Q^2 \geq 2$ GeV$^2$ (see reviews in \cite{CoDeRo}
and references therein), which gives us a reason to believe that
perturbative QCD is capable of describing the
evolution of the structure function (SF) $F_2$
% and its derivatives
%%structure functions
down to very low $Q^2$ values, where all the strong interactions
are conventionally considered to be soft processes.

Recently, the ZEUS and H1 Collaborations have presented
%several
the new precise combined data \cite{Aaron:2009aa} for the SF
%structure function
$F_2$.
In the recent paper \cite{Kotikov:2012sm}, we analyzed
%tried The aim of this short paper is to compare
the combined H1$\&$ZEUS data based on
%predictions %between
the predictions obtained by using the so-called generalized
doubled asymptotic scaling (DAS)
%generalized DAS
approach \cite{Q2evo}.

At ruther low $Q^2$ values, perturbation theory becomes to be marginal and
some resumation is very helpful.
Fortunately, at low $x$ values the magnitude of the NLO corrections is
strongly negative and effective increase of the scale of the strong coupling
constant reduces the  NLO corrections.
%With another side,
Moreover, the higher values
of the coupling constant scale leads to stabilization of prturbation theory
at lower $Q^2$ values.
Since in the framework of the generalized DAS approach \cite{Q2evo}, the
the singular and regular
parts of anomalous dimensions contribute to different parts of parton
distributions (PDs) it is natural to decrease (really even to neglect) the
singular part of the NLO anomalous dimansions.
Indeed,
%this
the singular parts contribute to the argument of the modified Bessel
functions and determinates the $Q^2$ evolution at low $Q^2$ values.

The aim of this short paper is to continue the analysis in
\cite{Kotikov:2012sm} by setting the
new scale of the coupling constant, which eliminates
the singular part of NLO anomalous dimesnions.
%at the next-to-leading (NLO) order of approximation.

%\section{Generalized DAS approach} \indent

{\bf I}. In  the generalized DAS approach \cite{Q2evo}
%The flat initial condition (\ref{1}) corresponds to the case when
parton densities
%distributions
tend  to some constant value at $x \to 0$ and at some initial value $Q^2_0$.
%(\ref{1}).
The main ingredients of the results \cite{Q2evo}, are:
\begin{itemize}
\item
%{\bf 1.}
Both, the gluon and quark singlet densities are presented in terms of two
components ($"+"$ and $"-"$) which are obtained from the analytic
$Q^2$-dependent expressions of the corresponding ($"+"$ and $"-"$) PD
%parton distributions
moments.
%\footnote{Such an approach has been developed  \cite{Albino:2011si}
%recently also for the fragmentation function,
%whose first moments (ie mean multiplicities of quarks and gluons) were
%analyzed \cite{Bolzoni:2012ii}. The results are in good agreement with the
%experimental data (see contribution \cite{Bolzoni:2012cv} by Paolo Bolzoni
%to this Proceedings).}
\item
%{\bf 2.}
The twist-two part of the $"-"$ component is constant at small $x$ at any
values of $Q^2$,
whereas the one of the $"+"$ component grows at $Q^2 \geq Q^2_0$ as
\begin{equation}
\sim e^{\sigma},~~~
%\exp{\sigma},~~~
\sigma = 2\sqrt{\left[ \left|\hat{d}_+\right| s
%\ln \left( \frac{a_s(Q^2_0)}{a_s(Q^2)} \right)
- \left( \hat{d}_{++}
%\hat{D}_+
+  \left|\hat{d}_+\right|
%\hat{d}_+
\frac{\beta_1}{\beta_0} \right) p
% \Bigl( a_s(Q^2_0) - a_s(Q^2) \Bigr)
\right] \ln \left( \frac{1}{x} \right)}  \ ,~~~ \rho=\frac{\sigma}{2\ln(1/x)} \ ,
\label{intro:1}
\end{equation}
where $\sigma$ and $\rho$
%$=\sigma/(2\ln(1/x))$
are the generalized Ball--Forte
variables,
\begin{equation}
s=\ln \left( \frac{a_s(Q^2_0)}{a_s(Q^2)} \right),~~
p= a_s(Q^2_0) - a_s(Q^2),~~~
\hat{d}_+ = - \frac{12}{\beta_0},~~~
%\hat{D}_+
\hat{d}_{++} =  \frac{412}{27\beta_0}.
\label{intro:1a}
\end{equation}
\end{itemize}
%$8[23 C_A - 26 C_F]T_Rf/(9\beta_0)$.
%%\item
%%The recently observed difference between small $x$ behavior of sea
%%quark and gluon densities at $Q^2=Q^2_0$ are incorporated (in [IKP])
%%by high-twist corrections to the above twist-two approximation.
%\end{itemize}
Hereafter we use the notation
$a_s=\alpha_s/(4\pi)$.
The first two coefficients of the QCD $\beta$-function in the $\ms$-scheme
are $\beta_0 = 11 -(2/3) f$
%$(11/3) C_A - (4/3) T_R f$
and $\beta_1 =  102 -(38/3) f$
%$(2/3)[17 C_A^2 - 10 C_A T_R f - 6 C_F T_R f]$
with $f$ is being the number of active quark flavors.
%This new presentation as a function of the
%$SU(N)$ group casimirs, with $f$ active flavors, $C_A = N$, $T_R = 1/2$,
%$T_F = T_R f$ and  $C_F = (N^2 - 1)/(2N)$ permits to apply our results
%to,  for example, the popular $N=1$ supersymmetric model.
%Of course, for $N=3$ one has the QCD result.
% \cite{Kotikov:1999}.
%
Note
%here
that the
%perturbative
coupling constant $a_s(Q^2)$ is different at
the leading-order (LO) and NLO approximations.
Usually at the NLO level ${\rm \overline{MS}}$-scheme is used, so we apply
$\Lambda = \Lambda_{\rm \overline{MS}}$ below.
%in the Eqs.~(\ref{an:NLO}) and (\ref{as:NLO}).

%\section{Grunberg approach}
{\bf II.} At low $Q^2$, the value of the strong coupling constant is large.
The most important terms, which are singular at $n \to 1$, where $n$
is the Mellin moment, are collected in the result (\ref{intro:1}) for
$\sigma^2$. So, it is convenient to choose the scale $\mu^2$ of the strong
coupland at which the
%most important
NLO contribution $\sim p$ in \
(\ref{intro:1}) vanishes. This choise is
\be
\mu^2 = Q^2 \exp\left[\left( \hat{d}_{++}
%\hat{D}_+
+ \left|\hat{d}_+\right|\beta_1/\beta_0 \right)/(\left|\hat{d}_+\right|\beta_0)
\right]
\approx 3.89 Q^2
\,
\label{mu2}
\ee
where the simbol $\approx$ marks the case
% Indeed,
%for the case
$f=3$, which is relevant at low $Q^2$ values. We see that
the choise (\ref{mu2}) increases effectively the argument of coupling constant
at low $x$ values (see \cite{DoShi}).
%, we have
%\be
%\mu^2 = 3 Q^2 \, .
%\label{mu2.1}
%\ee

\begin{figure}[t]
\centering
\vskip 0.5cm
\includegraphics[width=.95\hsize]{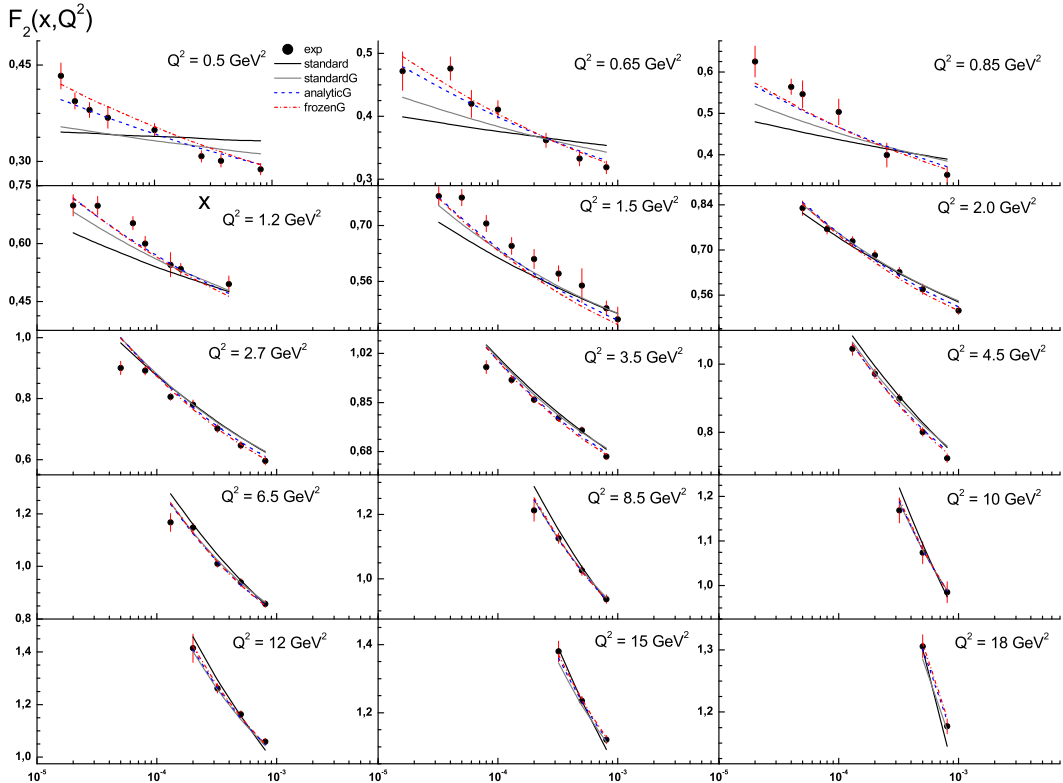}
\vskip -0.3cm
\caption{$x$ dependence of $F_2(x,Q^2)$ in bins of $Q^2$.
The combined experimental data from H1 and ZEUS Collaborations
\cite{Aaron:2009aa} are
compared with the NLO fits for $Q^2\geq0.5$~GeV$^2$ implemented with the
scales of the strong coupling constant equal to $Q^2$
%standard
(solid lines) and to $\mu^2$ (gray lines). The dot-dashed and dashed lines
mark, respectively, the results based on the frozen and analytic
%, frozen (dot-dashed lines), and analytic (dashed lines)
versions of the strong coupling constant.}
\label{fig:F1}
\end{figure}

%\section{``Frozen'' and analytic  coupling constants}

%{\bf IV.}
In order to improve an agreement at lowest $Q^2$ values,
the QCD coupling constant is modified in the infrared region.
We consider two modifications.
% that effectively increase the argument of the coupling constant
%at low $Q^2$ values (see \cite{DoShi}).
In the first case, which is more phenomenological, we introduce freezing
of the coupling constant by changing its argument $Q^2 \to Q^2 + M^2_{\rho}$,
where $M_{\rho}$ is the $\rho $-meson mass (see \cite{Cvetic1} and
%references
discussions therein). Thus, in the formulae (\ref{intro:1}) and
(\ref{intro:1a})
%of Sec. 2
we have to carry out the following replacement:
\begin{equation}
 a_s(Q^2) \to a_{\rm fr}(Q^2) \equiv a_s(Q^2 + M^2_{\rho})
\label{Intro:2}
\end{equation}

The second possibility follows the Shirkov--Solovtsov idea
\cite{ShiSo}
concerning the analyticity of the coupling constant that leads
to additional power dependence of the latter.
Then, in the formulae of the previous section
%and \ref{Sec:3}
the coupling constant $a_s(Q^2)$ should be replaced as follows:
\begin{eqnarray}
 a^{\rm LO}_{\rm an}(Q^2) \, = \, a^{\rm LO}_s(Q^2) - \frac{1}{\beta_0}
 \frac{\Lambda^2_{\rm LO}}{Q^2 - \Lambda^2_{\rm LO}},~~~~
 a_{\rm an}(Q^2) \, = \, a_s(Q^2) - \frac{1}{2\beta_0}
 \frac{\Lambda^2}{Q^2 - \Lambda^2}
+ \ldots \, ,
\label{an:LO}
\end{eqnarray}
in the LO and NLO approximations, respectively.
%and
%\begin{eqnarray}
% a_{\rm an}(Q^2) \, = \, a_s(Q^2) - \frac{1}{2\beta_0}
% \frac{\Lambda^2}{Q^2 - \Lambda^2}
%+ \ldots \, ,
%%- \frac{1}{\beta_0}
%% \sum_{k=1}^\infty \left(\frac{\Lambda^2}{Q^2}\right)^k \, C_k[f]
%\label{an:NLO}
%\end{eqnarray}
%in the NLO approximation.
Here the the symbol $\ldots$ stands
for the terms that provide negligible contributions when $Q^2 \geq 1$ GeV \cite{ShiSo}.

%\section{Comparison with experimental data} \indent

{\bf III.} By using the generalized DAS approach
%results of the previous section
we have
analyzed H1$\&$ZEUS data for $F_2$ \cite{Aaron:2009aa}.
% and the slope $\partial \ln F_2/\partial \ln (1/x)$ \cite{H1slo,DIS02}
%at small $x$ from the H1 and ZEUS Collaborations.
%
In order to keep the analysis as simple as possible,
we fix $f=4$ and $\alpha_s(M^2_Z)=0.1168$ (i.e., $\Lambda^{(4)} = 284$ MeV) in agreement
with
%more recent
ZEUS results given in~\cite{H1ZEUS}.

As can be seen from Fig.~1
%and Table~1,
%(see also \cite{Q2evo,HT}),
the twist-two approximation is reasonable for $Q^2 \geq 2$ GeV$^2$.
At lower $Q^2$, we observe that
the choise (\ref{mu2}) of the $\mu^2$ scale provides only a little improvement
in an agreement with data. However,
the fits in the cases with ``frozen'' and
analytic strong coupling constants, which  are very close each other
%similar
(see also \cite{KoLiZo,Cvetic1}), describe the data in
the low $Q^2$ region significantly better than the standard fits
wiht $Q^2$ and $\mu^2$ scales.
Nevertheless, for $Q^2 \leq 1.5$~GeV$^2$
%in the case of $\lambda^{\rm eff}_{F_2}(x,Q^2_0)$,
there is still some disagreement with
the data, which needs to be additionally studied.
In particular,  the Balitsky--Fadin--Kuraev--Lipatov
%(BFKL)
resummation \cite{BFKL} may be important here \cite{Kowalski:2012ur}.
It can be added in the generalized DAS approach according to the discussion
in Ref. \cite{KoBaldin}.

%\vspace{-0.3cm}

%\section*{Acknowledgements}
{\bf Acknowledgements}
A.V.K. thanks the Organizing Committee of International Workshop
``Physics of Fundamental Interactions'' for invitation and support.
This work was
supported in part by RFBR grant 13-02-01005-a.

%\vspace{-0.3cm}

\end{document}